\documentclass{wscpaperproc}
\usepackage{latexsym}
\usepackage{graphicx}
\usepackage{mathptmx}

\usepackage{amsmath}
\usepackage{amsfonts}
\usepackage{amssymb}
\usepackage{amsbsy}
\usepackage{amsthm}
\usepackage{subcaption}
\usepackage{float}

\usepackage[pdftex,colorlinks=true,urlcolor=blue,citecolor=black,anchorcolor=black,linkcolor=black]{hyperref}

\newtheoremstyle{wsc}
{3pt}
{3pt}
{}
{}
{\bf}
{}
{.5em}
{}

\theoremstyle{wsc}

\begin{document}

%
%

\pagestyle{fancyplain}

\thispagestyle{plain}
\firstPageHead{}

\chead{\fancyplain{}{\itshape Van Steenkiste, van der Herten, Couckuyt, and Dhaene}}

\rhead{}
\cfoot{}
\renewcommand{\headrulewidth}{0pt} 

\makeatletter
\let\@internalcite\cite
\def\cite{\def\@citeseppen{-1000}%
    \def\@cite##1##2{(##1\if@tempswa , ##2\fi)}%
    \def\citeauthoryear##1##2##3{##1 ##3}\@internalcite}
\def\citeNP{\def\@citeseppen{-1000}%
    \def\@cite##1##2{##1\if@tempswa , ##2\fi}%
    \def\citeauthoryear##1##2##3{##1 ##3}\@internalcite}
\def\citeN{\def\@citeseppen{-1000}%
    \def\@cite##1##2{##1\if@tempswa, ##2)\else{}\fi}%
    \def\citeauthoryear##1##2##3{##1 (##3)}\@citedata}
\def\citeA{\def\@citeseppen{-1000}%
    \def\@cite##1##2{(##1\if@tempswa , ##2\fi)}%
    \def\citeauthoryear##1##2##3{##1}\@internalcite}
\def\citeANP{\def\@citeseppen{-1000}%
    \def\@cite##1##2{##1\if@tempswa , ##2\fi}%
    \def\citeauthoryear##1##2##3{##1}\@internalcite}
\def\shortcite{\def\@citeseppen{-1000}%
    \def\@cite##1##2{(##1\if@tempswa , ##2\fi)}%
    \def\citeauthoryear##1##2##3{##2 ##3}\@internalcite}
\def\shortciteNP{\def\@citeseppen{-1000}%
    \def\@cite##1##2{##1\if@tempswa , ##2\fi}%
    \def\citeauthoryear##1##2##3{##2 ##3}\@internalcite}
\def\shortciteN{\def\@citeseppen{-1000}%
    \def\@cite##1##2{##1\if@tempswa, ##2\else{}\fi}%
    \def\citeauthoryear##1##2##3{##2 (##3)}\@citedata}
\def\shortciteA{\def\@citeseppen{-1000}%
    \def\@cite##1##2{(##1\if@tempswa , ##2\fi)}%
    \def\citeauthoryear##1##2##3{##2}\@internalcite}
\def\shortciteANP{\def\@citeseppen{-1000}%
    \def\@cite##1##2{##1\if@tempswa , ##2\fi}%
    \def\citeauthoryear##1##2##3{##2}\@internalcite}
\def\citeyear{\def\@citeseppen{-1000}%
    \def\@cite##1##2{(##1\if@tempswa , ##2\fi)}%
    \def\citeauthoryear##1##2##3{##3}\@citedata}
\def\citeyearNP{\def\@citeseppen{-1000}%
    \def\@cite##1##2{##1\if@tempswa , ##2\fi}%
    \def\citeauthoryear##1##2##3{##3}\@citedata}
%
%
%
\def\@citedata{%
    \@ifnextchar [{\@tempswatrue\@citedatax}%
                  {\@tempswafalse\@citedatax[]}%
}

\def\@citedatax[#1]#2{%
\if@filesw\immediate\write\@auxout{\string\citation{#2}}\fi%
  \def\@citea{}\@cite{\@for\@citeb:=#2\do%
    {\@citea\def\@citea{, }\@ifundefined
       {b@\@citeb}{{\bf ?}%
       \@warning{Citation `\@citeb' on page \thepage \space undefined}}%
{\csname b@\@citeb\endcsname}}}{#1}}%

%
\def\@citex[#1]#2{%
\if@filesw\immediate\write\@auxout{\string\citation{#2}}\fi%
  \def\@citea{}\@cite{\@for\@citeb:=#2\do%
    {\@citea\def\@citea{, }\@ifundefined
       {b@\@citeb}{{\bf ?}%
       \@warning{Citation `\@citeb' on page \thepage \space undefined}}%
{\csname b@\@citeb\endcsname}}}{#1}}%

%
\def\@biblabel#1{}
\makeatother



\newdimen\bibindent
\bibindent=0.0em
\def\thebibliography#1{\section*{\refname}\list
   {}{\settowidth\labelwidth{[#1]}
   \leftmargin\parindent
   \itemindent -\parindent
   \listparindent \itemindent
   \itemsep 0pt
   \parsep 0pt}
   \def\newblock{}
   \sloppy
   \sfcode`\.=1000\relax}


\setlength{\baselineskip}{12.7pt}

\title{SENSITIVITY ANALYSIS OF EXPENSIVE BLACK-BOX SYSTEMS USING METAMODELING}

\author{Tom Van Steenkiste\\ 
Joachim van der Herten\\
Ivo Couckuyt\\
Tom Dhaene\\[12pt]
Department of Information Technology \\
Ghent University\\
Technologiepark Zwijnaarde 15\\
Ghent, 9052, BELGIUM}

\maketitle

\section*{ABSTRACT}
Simulations are becoming ever more common as a tool for designing complex products. Sensitivity analysis techniques can be applied to these simulations to gain insight, or to reduce the complexity of the problem at hand. However, these simulators are often expensive to evaluate and sensitivity analysis typically requires a large amount of evaluations. Metamodeling has been successfully applied in the past to reduce the amount of required evaluations for design tasks such as optimization and design space exploration. In this paper, we propose a novel sensitivity analysis algorithm for variance and derivative based indices using sequential sampling and metamodeling. Several stopping criteria are proposed and investigated to keep the total number of evaluations minimal. The results show that both variance and derivative based techniques can be accurately computed with a minimal amount of evaluations using fast metamodels and FLOLA-Voronoi or density sequential sampling algorithms.

\section{INTRODUCTION}
\label{sec:introduction}

Simulations are a valuable tool in the design and analysis of complex problems. They offer researchers and engineers a better understanding of a problem without numerous expensive real-life experiments or prototypes. However, the increase of simulation accuracy over the years has significantly increased their evaluation time and computational requirements. A popular method is screening of the input variables using sensitivity analysis to reduce the complexity in subsequent analyses of the system. This provides information on how changes in the input affect the output \shortcite{Saltelli2004}. Within sensitivity analysis, several techniques can be defined, such as variance based methods and derivative based methods.

Although sensitivity analysis can be used to reduce complexity, the sensitivity analysis methods are often expensive themselves in terms of number of evaluations. Building a metamodel, also referred to as a surrogate model or a response surface model, of the simulator is a popular method for the analysis of the simulator output using a minimal amount of evaluations. Metamodeling techniques are popular for tasks such as optimization, reliability analysis as well as sensitivity analysis. The latter includes variance  based methods \shortcite{Jin2004} and derivative based methods \shortcite{Sudret2015}. In this paper, we explore variance and derivative based methods for the following metamodel types: Kriging \shortcite{Santner2003,Forrester2008}, Gaussian processes (GP) \shortcite{Rasmussen2006} and Least-Squares Support Vector Machines (LS-SVM) \shortcite{Suykens2002}. The traditional method to compute these indices uses Monte Carlo or quasi-Monte Carlo methods \shortcite{Sobol2001,Saltelli2002b}. However, for metamodels of a tensor product functional form, analytic derivations exist for variance based sensitivity indices \shortcite{Jin2004}. Furthermore, we derive an analytic formula for the derivative based sensitivity indices for such functions, similar to the variance based formula. Although Monte-Carlo methods can be applied to cheap metamodels, these analytic derivations offer improved accuracy and scalability for problems of higher dimensionality.

\begin{figure}[H]
        \includegraphics[width=\textwidth]{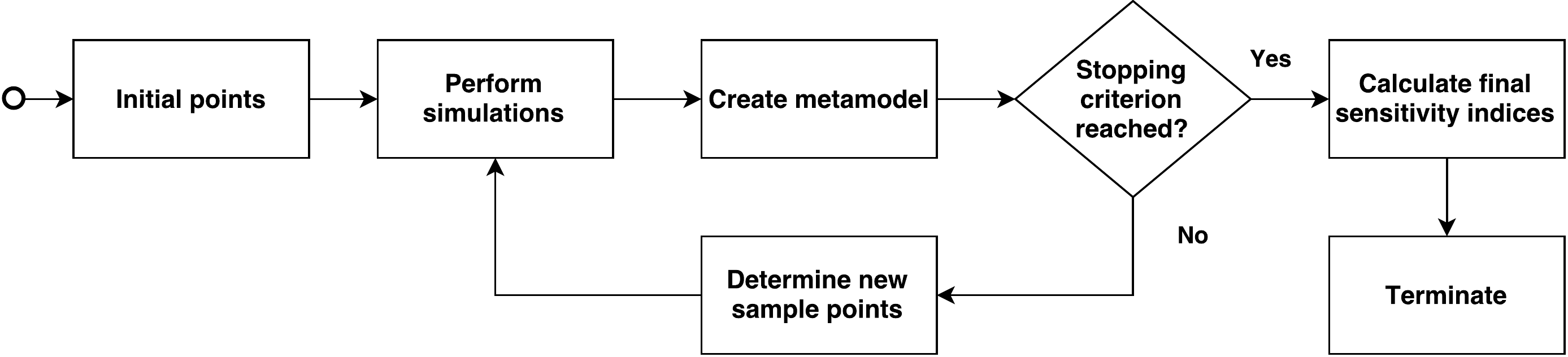}
        \caption{The metamodeling process with sequential design for sensitivity analysis.}
        \label{fig:metamodelingprocess}
\end{figure}

A metamodel can be built on a single pre-generated data set, after which the indices are calculated. However, determining the number and positions of data points corresponding to the best accuracy, is a difficult task. The performance and accuracy of metamodeling techniques are further improved when using sequential sampling techniques. Examples of which are FLOLA-Voronoi \shortcite{Herten2015} and density-based \shortcite{Crombecq2010a} designs. Starting from a very small initial number of data points, sequential sampling techniques use specialized criteria to generate additional data during the metamodeling process. In this paper, we use this sequential design methodology to efficiently compute sensitivity indices using metamodels. This results in a novel, efficient workflow for sensitivity analysis of computationally expensive simulations, which is illustrated in Figure~\ref{fig:metamodelingprocess}.

To complete this workflow, we explore different stopping criteria to determine when the indices are sufficiently accurate. A common stopping criterion used in metamodeling is the metamodel accuracy assessed with cross-validation using a specific error measure such as the Root-Relative-Square-Error (RRSE) or Bayesian-Estimation-Error-Quotient (BEEQ) \shortcite{Li2006} with equations 

\begin{align*}
\text{RRSE}(y,\tilde{y}) = \sqrt{\frac{\sum_{i=1}^n(y_i-\tilde{y}_i)^2}{\sum_{i=1}^n(y_i-\bar{y})^2}}, && \text{BEEQ}(y,\tilde{y}) = \Bigg( \prod_{i=1}^n \frac{|y_i-\tilde{y}_i|}{|y_i-\bar{y}|}  \Bigg)^{\frac{1}{n}}.
\end{align*}

In this paper, we propose a stopping criterion using cross-validation with two different measures directly based on the variance based and the derivative based sensitivity indices, to terminate the process when the sensitivity analysis results are sufficiently accurate. This measure can be configured to be the maximal or mean variance of the sensitivity indices across the folds. We investigate the performance and accuracy of traditional cross-validation on the metamodel accuracy, and compare it to the proposed measure of cross-validation on the sensitivity indices which is more interpretable by the analyst.

Metamodeling with Kriging, GP and LS-SVM using sequential sampling is explained in Section~\ref{sec:metamodeling}. In Section~\ref{sec:sensitivityanalysis}, a short introduction is given on variance and derivative based sensitivity analysis, followed by a description of the combination of sensitivity analysis with metamodeling techniques in Section~\ref{sec:metamodelingbasedsensitivityanalysis}. The experimental setup is detailed in Section~\ref{sec:experimentalsetup}. In Section~\ref{sec:resultsanddiscussion}, the results are presented and discussed. Finally, conclusions are made in Section~\ref{sec:conclusion}.

\section{METAMODELING}
\label{sec:metamodeling}
Metamodeling, also known as surrogate modeling, is a commonly used technique to analyze problems involving expensive simulators. A metamodel is fit on high quality data provided by evaluations of a black box simulator. Many improvements can be made to the basic metamodeling process, such as using sequential sampling to determine where to sample next and a stopping criterion to determine when to stop.
\subsection{Metamodels}
\label{subsec:models}
Many different models can be used for metamodeling. Among the most popular ones are Kriging \shortcite{Santner2003,Forrester2008}, GP \shortcite{Rasmussen2006} and LS-SVM \shortcite{Suykens2002}. Kriging is similar to a Gaussian process. Namely, it is a Gaussian process conditioned to interpolate the data. It is often used for various metamodeling purposes such as stochastic metamodeling \shortcite{Chen2013} or sensitivity analysis \shortcite{Wang2007}. The above metamodels all belong to the class of kernel based methods which has the form of 

$$\hat{f}(x) = \sum_{i=1}^N \alpha_i\,k(x,x_i),$$

\noindent where $N$ is the amount of basis vectors. When the kernel $k$ is separable, the formula can be written as 

\begin{equation} \label{eq:separablemodel}
\hat{f}(x) = \sum_{i=1}^N \alpha_i \prod_{l=1}^d h_{i,l}(x_l),
\end{equation}

\noindent where $h_{i,l}(x_l)$ is the part of the $d$-dimensional kernel $k$ for dimension $l$. Such a representation is known as a tensor-product function \shortcite{Jin2004}. A  popular kernel is the RBF kernel, $k(x,x_i)=\prod_{l=1}^d\exp(-\theta_i||x_l-x_{i,l}||^2)$, also known as the Gaussian or the squared exponential kernel. While the Mat\'{e}rn $\frac{3}{2}$ kernel is not separable we can define a separable version as the product of $d$ 1-dimensional Mat\'{e}rn $\frac{3}{2}$ kernels

$$k(x,x_i)=\prod_{l=1}^d(1+\sqrt{3}\theta_i||x_l-x_{i,l}||)\exp(-\sqrt{3}\theta_i||x_l-x_{i,l}||).$$

The metamodels can be trained with an individual hyperparamter $\theta_i$ for each dimension or a single hyperparameter $\theta$ for all dimensions. Details of the workings of these metamodels are discussed in the referenced literature.

\subsection{Sequential Design}
\label{subsec:sequentialdesign}
In a basic metamodeling setup, the number and location of sample points for simulator evaluations is determined up front. This is traditionally done with one-shot approaches such as factorial designs \shortcite{Lehmensiek2002}, optimized Latin hypercubes \shortcite{Dam2009}, etc. However, one-shot designs have the risk of under-fitting (too few data) or over-fitting (too much data). A sequential sampling approach can be used to improve metamodeling efficiency and accuracy. Instead of the up-front one-shot design, a small set of initial data points is iteratively extended with additional samples. Each iteration an intermediate metamodel is built and analyzed together with the simulator responses. Doing so, sequential design  can exploit the available information (exploitation) in addition to space-filling criteria (exploration) to modify the distribution of the samples to the problem and application at hand. Additionally, the process can be halted when the predefined goals such as metamodel accuracy have been met, reducing the amount of simulator evaluations.

An example of a sequential design strategy is FLOLA-Voronoi, introduced in \shortcite{Herten2015}. It is a computationally efficient approach for increasing the amount of samples in non-linear regions which is beneficent as these non-linear regions are more difficult to model. The FLOLA-Voronoi algorithm combines a gradient estimation in the datapoints, based on a locally linear approximation, for exploitation, with a Voronoi space-filling criterion for exploration \shortcite{Herten2015}. FLOLA-Voronoi offers a fuzzy approach for determining the neighbors in the exploitation step, increasing the efficiency in higher dimensions over regular LOLA-Voronoi \shortcite{Crombecq2010a}.

Another sequential sampling technique is the density based design, introduced in \shortcite{Crombecq2011}. In this method, points are generated taking into account the maximin distance and the projected distance. This projected distance property is particularly useful for screening purposes.

\section{SENSITIVITY ANALYSIS}
\label{sec:sensitivityanalysis}

Global sensitivity analysis is an important approach in simulation to determine how the output behavior is related to changes in the inputs \shortcite{Saltelli2002}. Various techniques for global sensitivity analysis exist, such as the variance based methods and the derivative based methods.

In variance based global sensitivity analysis, the variance in the input is related to the variance in the output. The $d$-dimensional function to be analyzed is decomposed using an ANOVA decomposition and the variance of the function is specified as a combination of the variances of the decomposed parts \shortcite{Sobol2001} according to

$$f(x_1,x_2,...,x_d) = f_0 + \sum_{i=1}^d f_i(x_i) + \sum_{i_1=1}^{d} \sum_{i_2=i_1+1}^{d} f_{i_1i_2}(x_{i_1},x_{i_2}) + ... + f_{1..d}(x_1,...,x_d),$$

$$V = \sum_{i=1}^d V_i + \sum_{i_1=1}^{d} \sum_{i_2=i_1+1}^{d} V_{i_1i_2} + ... + V_{1..d}.$$

The Sobol sensitivity index for a subset $U$ of inputs is then calculated as $S_{U}=V_{U}/V$ \shortcite{Sobol2001}. This is the variance attributed to those inputs, normalized by the total variance of the function. The sensitivity indices are classified as either main effects when only one index is chosen, or interaction effects when multiple indices are chosen. A total sensitivity index $S_{i}^T$ is defined as the sum of the main effect and all interaction effects containing a specific input $i$.

Variance based global sensitivity indices, as defined above, are easily interpretable by the analyst. Unfortunately, they generally require a large amount of function evaluations \shortcite{Kucherenko2009,Touzani2014}. Using a metamodeling technique, the amount of evaluations can be greatly reduced. However, variance based sensitivity indices use the general assumption that variance is sufficient to describe output variance \shortcite{Saltelli2002}. To improve the analysis of the problem, we also investigate another index type called the derivative based global sensitivity index (DGSM). This index has evolved from the elementary effects method by \shortciteN{Morris1991} into several different definitions. In this paper, we use a recent definition by \shortciteN{Sobol2009}

\begin{equation} \label{eq:dgsm}
\nu_i = \mathop{\mathbb{E}}\bigg[\Big(\frac{\partial f}{\partial x_i}(x_1,...,x_d)\Big)^2\bigg],
\end{equation}

\noindent where the domain of the function to be analyzed is the unit hypercube $\mathcal{H}^d$. Although these derivative based sensitivity indices are less interpretable than the variance based indices, they can be linked to the total Sobol indices as an upper bound using the definition in \shortcite{Sobol2009,Lamboni2013}

\begin{equation} \label{eq:dgsmvstotalsobol}
S_i^T \leq S_i^{DGSM} = \frac{\nu_i}{\pi^2V},
\end{equation}

\noindent where $V$ is the variance of the function.

Many different definitions for sensitivity analysis indices have been proposed to overcome some of the deficiencies mentioned of the above indices. Examples are the non-moment based sensitivity indices by \shortciteN{Borgonovo2007} that overcome the projection of a distribution on a single moment or the interpretable derivative based indices by \shortciteN{Touzani2014} that overcome the interpretation issues. These can all be calculated based on metamodels using, e.g., Monte Carlo methods. However, the focus of this work is to analytically derive the sensitivity indices from the metamodel, which is much more efficient for higher dimensions. Hence, we limit ourselves to the measures discussed above which also have a much wider presence and are commonly used by analysts.

\section{METAMODELING BASED SENSITIVITY ANALYSIS}
\label{sec:metamodelingbasedsensitivityanalysis}
In this section, sensitivity analysis is incorporated into a metamodeling process. First we provide analytic derivations of the sensitivity indices for the described metamodels in \ref{subsec:variancebased} and~\ref{subsec:derivativebased}. Then, we define a stopping criterion to be used together with the sequential sampling strategies from Section~\ref{sec:metamodeling}, in order to provide a complete algorithm for accurately determining the sensitivity indices with a minimal amount of simulator evaluations.

\subsection{Variance Based}
\label{subsec:variancebased}
For metamodels that can be written as a tensor-product functional form (see Section~\ref{sec:metamodeling}), the Sobol indices can be derived analytically \shortcite{Jin2004}. For such metamodels, defined by the $\alpha_i$ and $h_{i,l}(x_l)$ functions in Equation~(\ref{eq:separablemodel}), the variance of a subset $V_{U}$ of input dimensions $U$ with $p_l(x)$ the distribution of the input $l$, $d$ the number of input dimensions and $N$ the number of samples can be computed using

$$ V_U = \sum_{i_1}^N \sum_{i_2}^N \Big( \alpha_{i_1}\alpha_{i_2}  \prod_{l=1}^d(C1_{i1,l}\;C1_{i2,l})(\prod_{l \in U} \frac{C2_{i_1,i_2,l}}{C1_{i1,l}\;C1_{i2,l}}-1)\Big),$$

$$C1_{i,l} = \int h_{i,l}(x_l)p_l(x_l)dx_l,$$

$$C2_{i_1,i_2,l} = \int h_{i_1,l}(x_l)h_{i_2,l}(x_l)p_l(x_l)dx_l.$$

For a complete derivation of these formulas we refer to the original works of \shortciteN{Jin2004}. By taking different subsets $U$ we can determine the total variance of the simulator and the variance of the main and interaction effects leading to all requirements to determine the Sobol and total Sobol indices for each input. By using these analytic formulas, we avoid the use of Monte Carlo methods directly on the simulator or the metamodel which would introduce additional errors and is infeasible for high-dimensional problems.

\subsection{Derivative Based}
\label{subsec:derivativebased}
Similarly, for the derivative based sensitivity indices, an analytic form of Equation~(\ref{eq:dgsm}) for derivative based indices can also be extracted for specific metamodels. This has been applied to polynomial chaos expansion metamodels in \shortciteN{Sudret2015}. Here, we derive the formulas for metamodels of the tensor-product form (see Section ~\ref{sec:metamodeling}). This results in the following equations

$$\nu_i = \sum_{i_1}^N\sum_{i_2}^N \alpha_{i_1}\alpha_{i_2} (\prod_{\substack{l=1 \\ l \neq i}}^d C2_{i_1,i_2,l})~C3_{i_1,i_2,i},$$

$$C3_{i_1,i_2,i} = \int \frac{\partial h_{i,i_1}(x_i)}{\partial x_i}\frac{\partial h_{i,i_2}(x_i)}{\partial x_i}p_i(x_i)dx_i.$$

The factor $C2$ in this equation is the same factor as for the variance based sensitivity indices. Because of this, both types of indices can be determined with a minimal amount of extra computations as only the $C3$ factor needs to be determined. Note that for the calculation of the $C3$ factor, the metamodel should support the calculation of the derivatives, which is available for many methods. If not, the derivation is typically straightforward. 

\subsection{Stopping Criterion}
\label{subsec:stoppingcriterion}
To avoid doing unnecessary extra simulator evaluations, a stopping criterion is defined to indicate the accuracy of the sensitivity indices. While error cross-validation can be used to assess the accuracy of the metamodel itself, it is potentially more interesting to directly calculate the cross-validation score of the sensitivity indices as this is more interpretable. The two following stopping criteria, based on cross-validation are proposed

\begin{align*}
\text{Mean}=\sum_{i=1}^d \frac{\mathrm{Var}(I_1,..,I_k)}{d}, && \text{Max}=\max_{i=1}^d \frac{\mathrm{Var}(I_1,..,I_k)}{d},
\end{align*}

\noindent where $k$ is the number of folds and $I_i$, a specific sensitivity index for variable $i$. The second criterion is stricter than the first one as it requires all sensitivity indices to be under the same predetermined threshold.

When the criterion reaches a predetermined threshold, the algorithm stops as the desired accuracy has been achieved which reduces the amount of expensive simulator evaluations needed. Depending on the application, a stricter or more relaxed threshold can be chosen. 

For derivative based indices, we propose a normalized version of these stopping criteria in which the indices are first normalized per evaluation, before computing the criterion. This form is proposed to overcome the interpretation problem discussed in Section~\ref{sec:sensitivityanalysis}.

\subsection{Complete Workflow}
\label{subsec:completeworklow}
The complete workflow of the algorithm is shown in Figure~\ref{fig:metamodelingprocess}. First an initial set of points is chosen using for example a Latin hypercube design after which the metamodeling loop starts. The chosen points are evaluated and a metamodel is built using all available points. The sensitivity indices are then evaluated analytically on the metamodel equation. At the end of each iteration, a stopping criterion is computed which can either be a model accuracy based error such as the RRSE or BEEQ or one of the criteria based on cross-validation with the sensitivity indices. If the stopping criterion threshold has been reached, the algorithm terminates. If it has not been reached, a new set of points are added using a sequential sampling strategy such as FLOLA-Voronoi or density based sampling after which another metamodeling iteration begins.
\section{EXPERIMENTAL SETUP}
\label{sec:experimentalsetup}
The performance and accuracy of the proposed global sensitivity analysis scheme is evaluated in an experiment on three typical sensitivity analysis benchmark functions and one real-world application.  Table~\ref{table:sensitivityanalysisfunctions} provides an overview of the mathematical formulation and dimensionality of the functions. For the Ishigami function, the domain of each input is $[-\pi,\pi]$ and for the G-Function and Moon-Function, the domain of each input is $[0,1]$.

\begin{table}[h]
\centering
\caption{Sensitivity analysis functions where $\epsilon$ collects the remaining, insignificant terms.}
\label{table:sensitivityanalysisfunctions}
\begin{tabular}{lll}
\hline
Name              & Equation & Inputs \\ \hline
\rule{0pt}{5ex}\shortstack[l]{\textbf{Ishigami}\\\shortcite{Ishigami1990}} & \(\displaystyle  \sin(x_1) + 7\sin(x_2)^2 + 0.1x_3^4\sin(x_1) \)         & 3            
\\
\rule{0pt}{5ex}\shortstack[l]{\textbf{G-function}\\\shortcite{Saltelli1994}}    & \(\displaystyle  \prod_{i=1}^3 \frac{|4x_i-2| + (i-2)/2}{1+(i-2)/2}  \)        & 3              \\
\shortstack[l]{\textbf{Moon}\\\shortcite{Moon2010}}  & \(\displaystyle  -19.71x_1x_{18}+23.72x_1x_{19}-13.34x_{19}^2+28.99x_7x_{12} + \epsilon \)       & 20            
\\
\hline
\end{tabular}
\end{table}

A real-world application is added to the experiment to test real-world applicability of the algorithm. The application is a braking system for satellites \shortcite{Kestila2013,Khurshid2014} called AaltoBrake. This simulator has a 5-dimensional input vector containing, among others, the mass of the braking system.

The experiments are performed using the SUMO-Toolbox \shortcite{Gorissen2010}. The Kriging model uses the implementation from ooDACE \shortcite{Couckuyt2013}, whereas the Gaussian process model uses the GPML library \shortcite{Rasmussen2010}. They are both optimized using maximum likelihood estimation. The Kriging and GP models both use a separable Mat\'{e}rn $\frac{3}{2}$ kernel and the LS-SVM model uses a Gaussian RBF kernel of which the hyperparameter is trained using RRSE cross-validation. For the low-dimensional problems ($<$ 20D), a hyperparameter is trained for each dimension of the Mat\'{e}rn kernel. For high-dimensional problems ($\geq$ 20D), a single hyperparameter is trained to avoid a complex and lengthy high-dimensional optimization of the hyperparameters. For the Gaussian RBF kernel, a single hyperparameter is trained regardless of dimensionality. 

The initial design used to build the metamodels is a Latin hypercube constructed using the Translational Propagation algorithm \shortcite{Viana2010}. During each step of the process, 10 new sampling points, determined by a FLOLA-Voronoi or density sequential design, are evaluated. A total of 300 samples are evaluated. This metamodeling process is shown in Figure~\ref{fig:metamodelingprocess}.

The complete metamodeling process is repeated 10 times for statistical robustness and at each step in the metamodeling process, several measures are calculated. The accuracy of the metamodel itself is evaluated each time using a 10-fold cross-validation with an RRSE measure and a BEEQ measure. Both the derivative based and the variance based sensitivity indices are calculated and their accuracy is evaluated using a 10-fold cross-validation setup with both the suggested stopping criteria from Section~\ref{sec:metamodelingbasedsensitivityanalysis}.

\section{RESULTS AND DISCUSSION}
\label{sec:resultsanddiscussion}
The results presented in this section are for the Kriging model with a FLOLA-Voronoi sequential design, unless otherwise specified. The figures show the mean and standard deviation across the 10 runs.

\begin{figure}[htb]
    \centering
    \begin{subfigure}[h]{0.48\textwidth}
        \includegraphics[width=\textwidth]{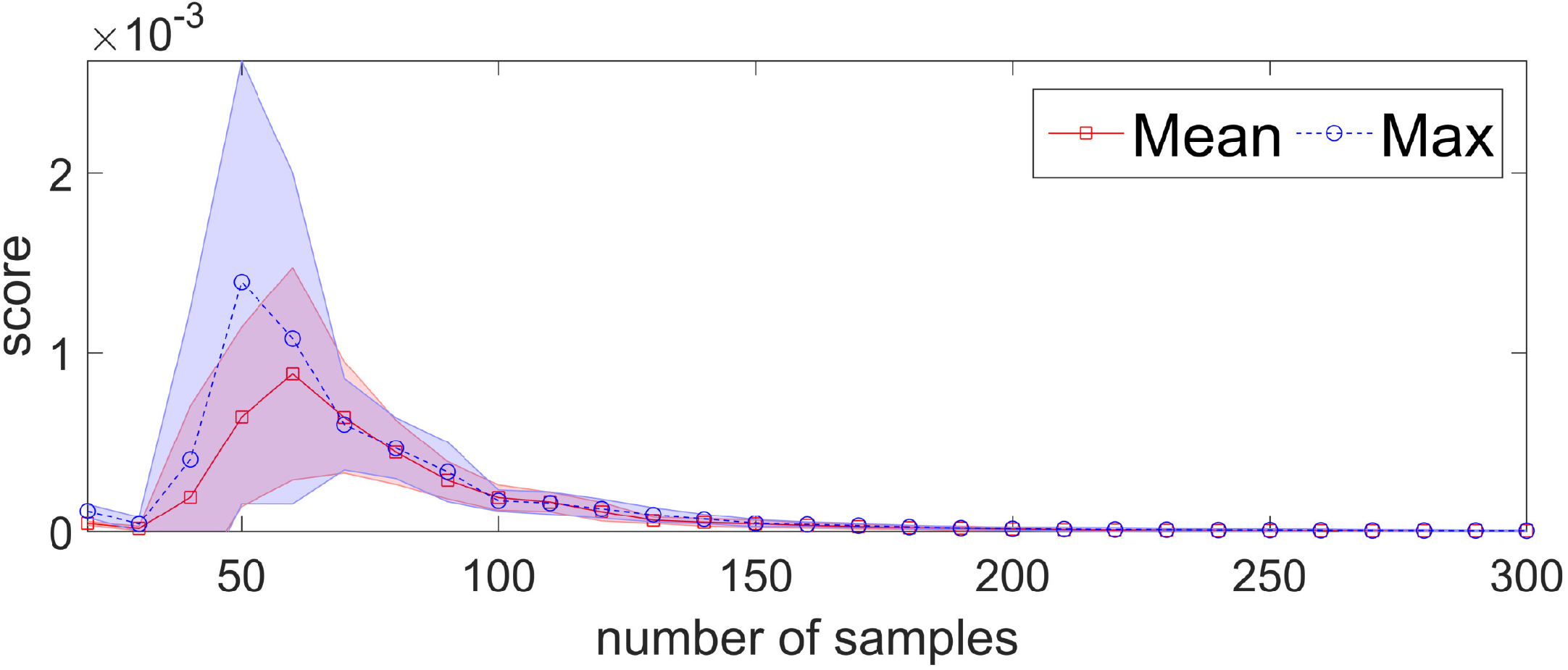}
        \caption{Ishigami.}
        \label{fig:ishigamisensitivycross-validation}
    \end{subfigure}
	~
    \begin{subfigure}[h]{0.48\textwidth}
        \includegraphics[width=\textwidth]{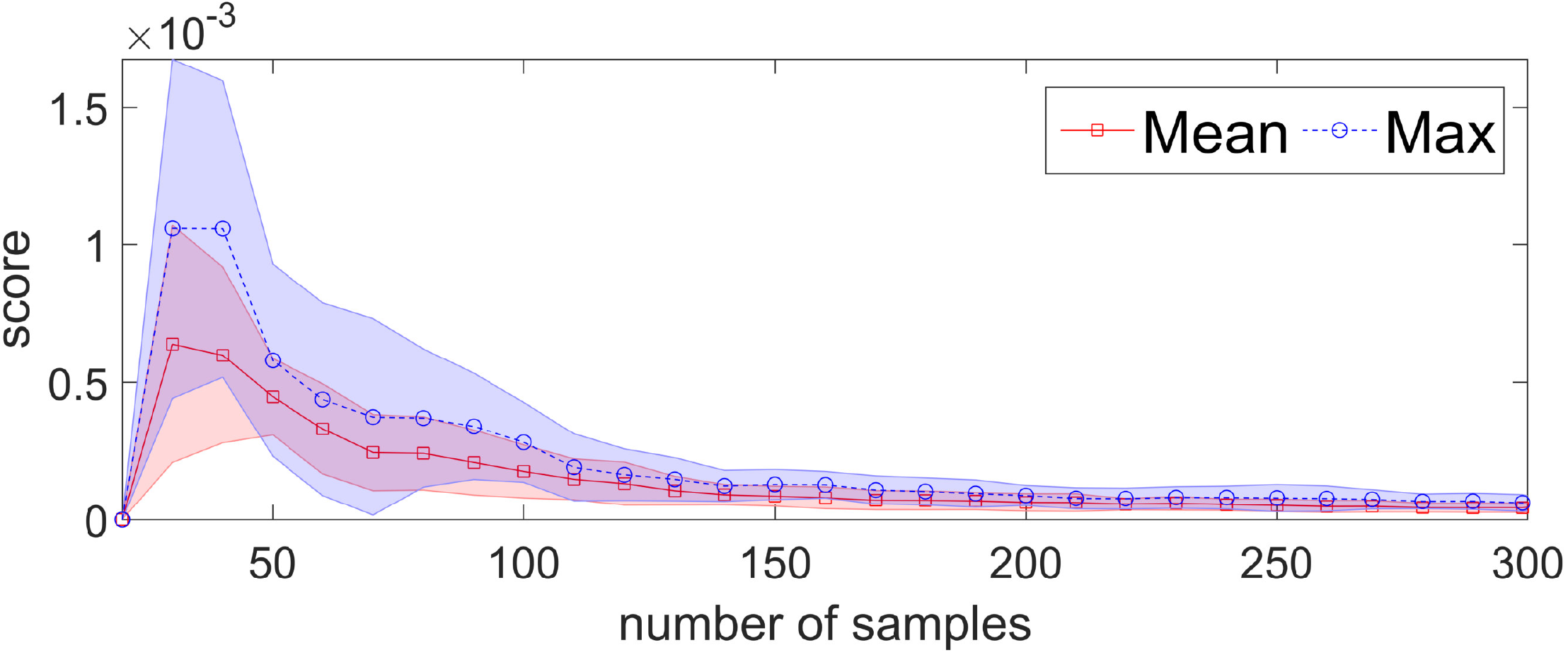}
        \caption{G-function.}
        \label{fig:gfunctionsensitivycross-validation}
    \end{subfigure}
    
    \begin{subfigure}[h]{0.48\textwidth}
        \includegraphics[width=\textwidth]{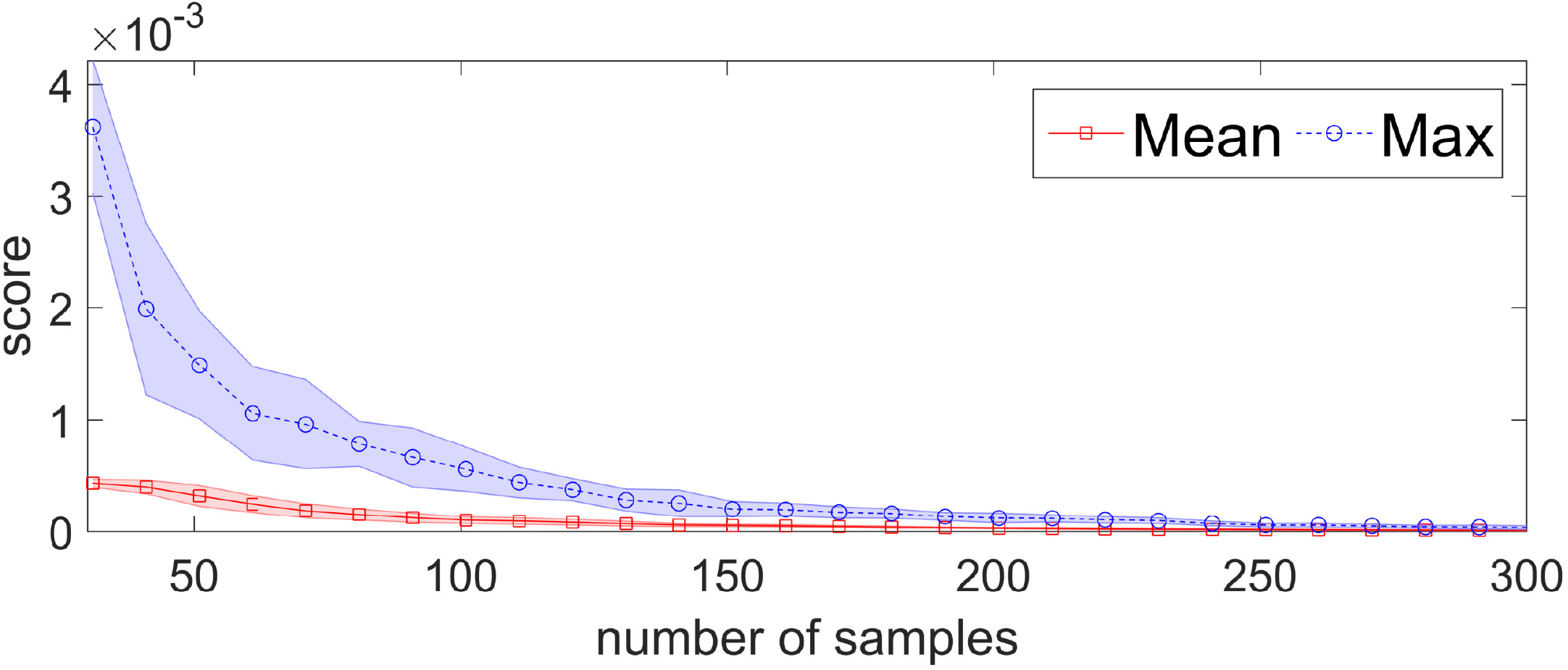}
        \caption{Moon.}
        \label{fig:moonsensitivycross-validation}
    \end{subfigure}
    ~
    \begin{subfigure}[h]{0.48\textwidth}
        \includegraphics[width=\textwidth]{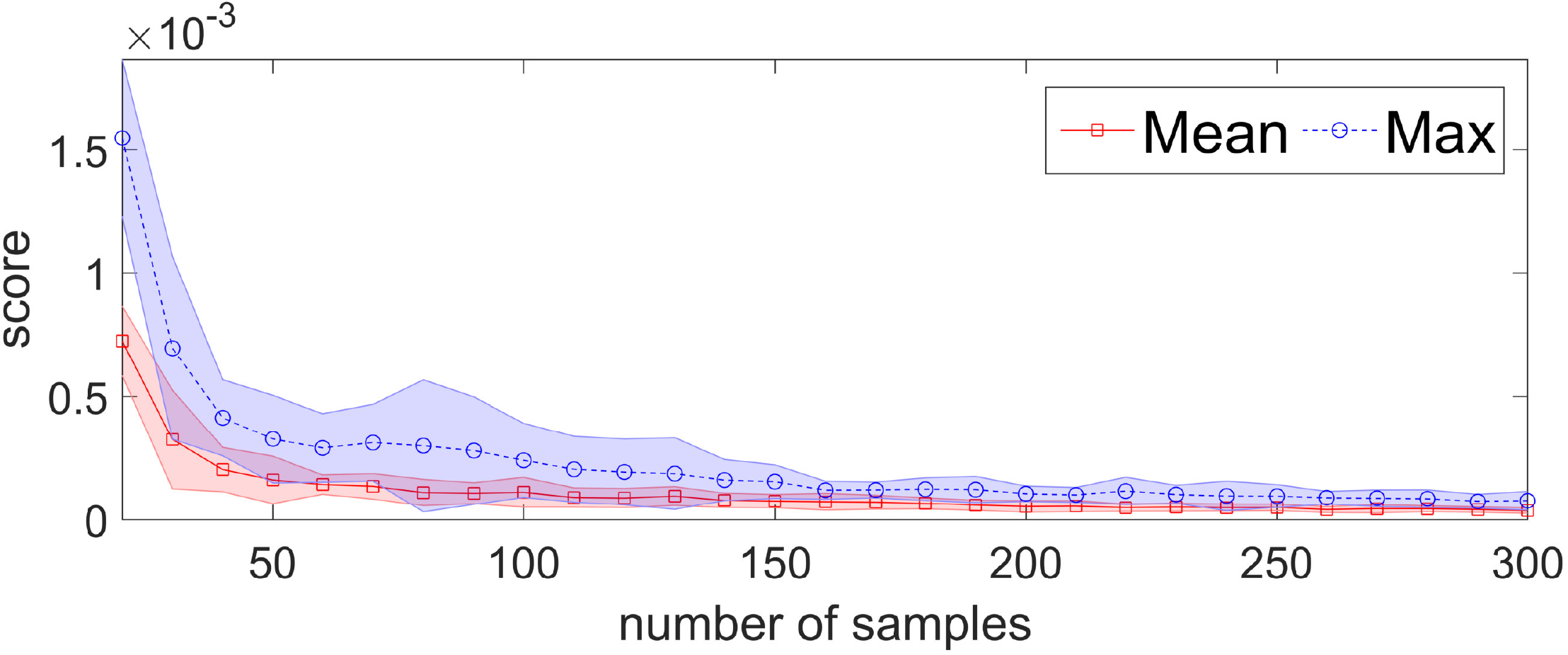}
        \caption{AaltoBrake.}
        \label{fig:aaltobreaksensitivycross-validation}
    \end{subfigure}

    \caption{Variance based sensitivity cross-validation measures for all examples with Kriging and FLOLA-Voronoi.}\label{fig:sensitivycross-validation}
\end{figure}

To evaluate the performance of the algorithm across all proposed experiments, the variance based sensitivity cross-validation measures for each experiment are shown in Figure~\ref{fig:sensitivycross-validation}.

There is no significant difference when using density sequential sampling, compared to using FLOLA-Voronoi for our test problems. For some cases in Figure~\ref{fig:sensitivycross-validation}, the proposed measures first increase. This is due to the metamodel struggling to fit the data accurately when there are few data available. This should be taken into account when checking the threshold.

\begin{figure}[htb]
	\begin{subfigure}[h]{0.48\textwidth}
        \includegraphics[width=\textwidth]{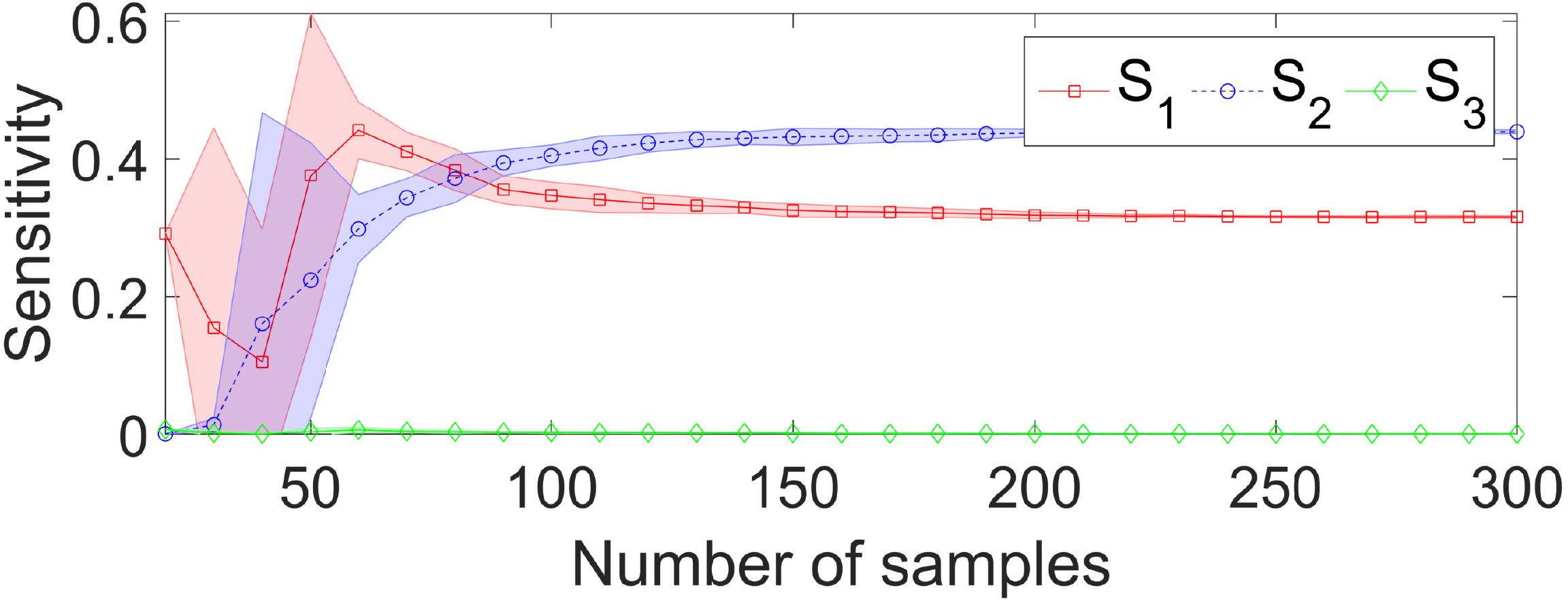}
        \caption{Sensitivity indices.}
        \label{fig:ishigamicomparisonindices}
    \end{subfigure}
    ~
	\begin{subfigure}[h]{0.48\textwidth}
        \includegraphics[width=\textwidth]{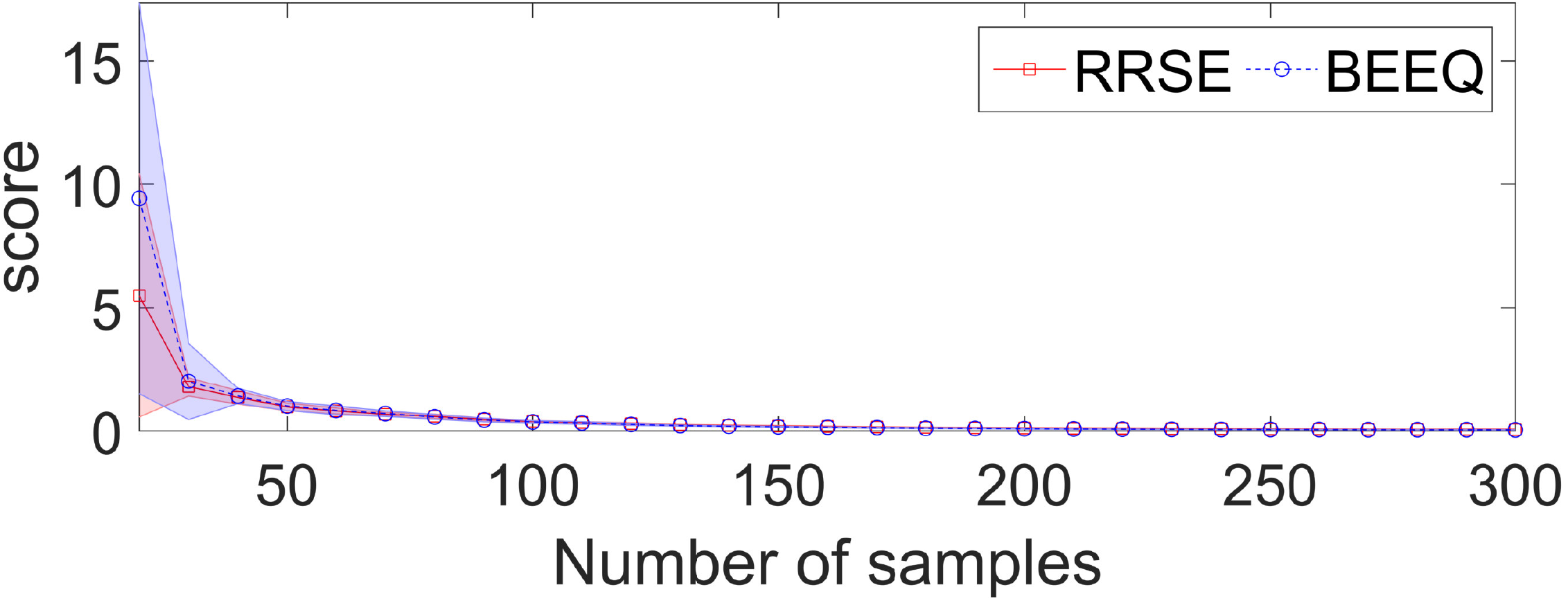}
        \caption{Error cross-validation.}
        \label{fig:ishigamicomparisonerrorcross-validation}
    \end{subfigure}
    \caption{Error cross-validation and variance based sensitivity indices for Ishigami with Kriging and FLOLA-Voronoi.}\label{fig:ishigamicomparison}
\end{figure}

Figure~\ref{fig:ishigamicomparisonindices} shows the variance based sensitivity indices for the Ishigami function and Figure~\ref{fig:ishigamicomparisonerrorcross-validation} shows the cross-validation measures based on RRSE and BEEQ. The accuracy based cross-validation measures are insufficient to use as a stopping criterion. When comparing Figure~\ref{fig:ishigamicomparisonerrorcross-validation} with Figure~\ref{fig:ishigamisensitivycross-validation}, both RRSE and BEEQ are too optimistic and have dropped significantly before the sensitivity indices have stabilized. The sensitivity cross-validation measure however, accurately reflects the time at which these indices stabilize.

\begin{figure}[H]
    \centering
    \begin{subfigure}[h]{0.48\textwidth}
        \includegraphics[width=\textwidth]{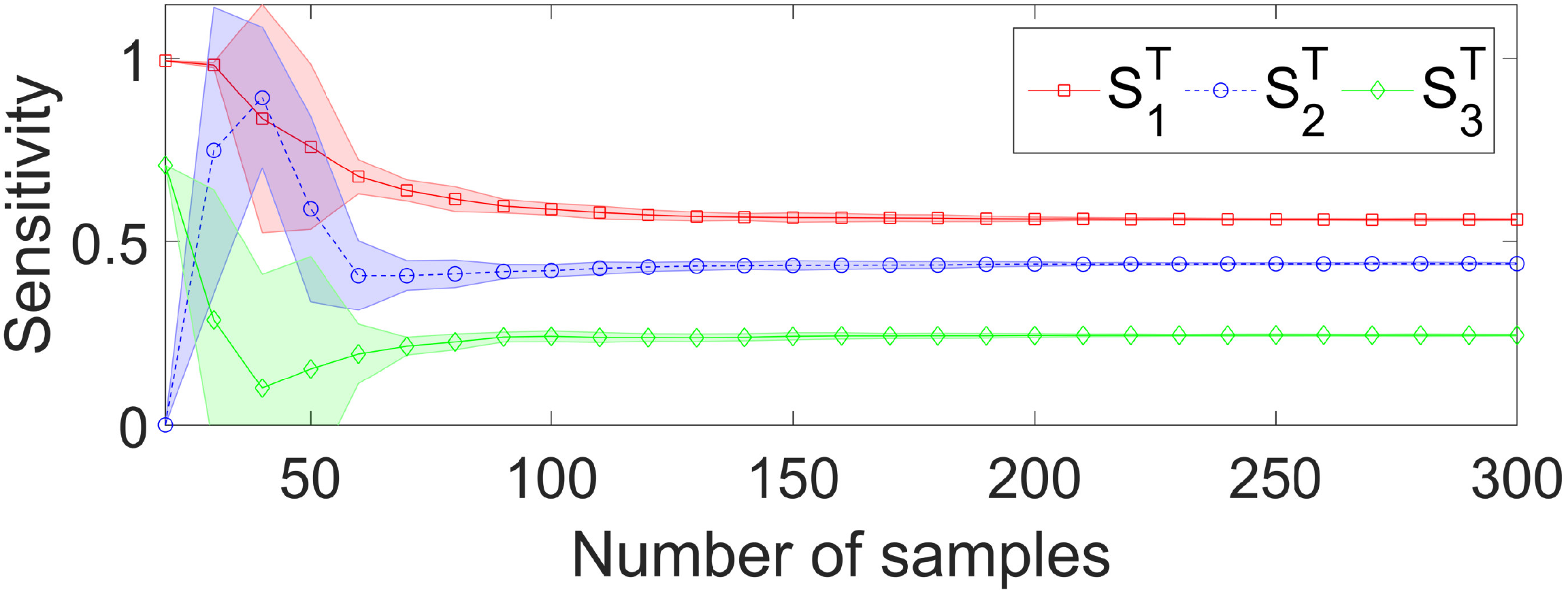}
        \caption{Total sensitivity indices.}
        \label{fig:ishigamitotalindices}
    \end{subfigure}
    ~
    \begin{subfigure}[h]{0.48\textwidth}
        \includegraphics[width=\textwidth]{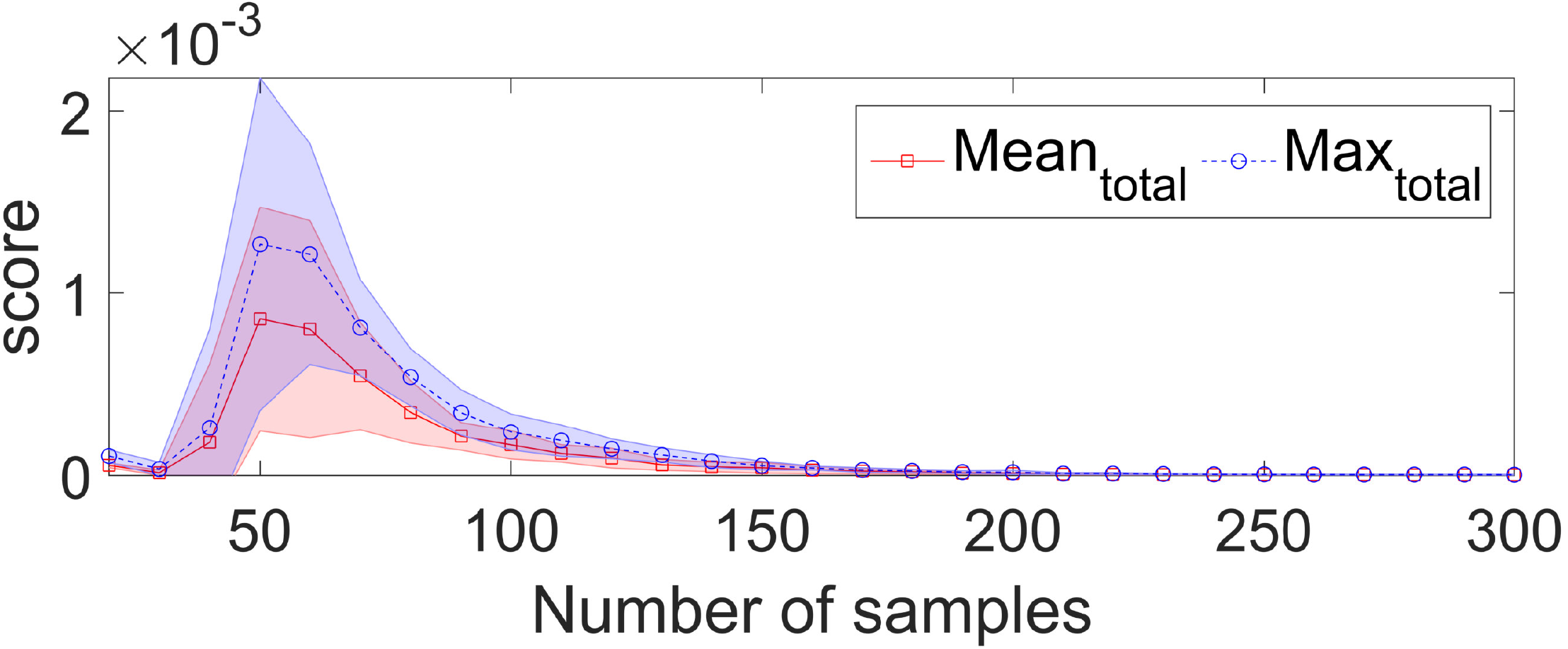}
        \caption{Total sensitivity cross-validation.}
        \label{fig:ishigamisensitivycross-validationvariancetotal}
    \end{subfigure}
    \caption{Variance-based total sensitivity cross-validation with corresponding total sensitivity indices for Ishigami with Kriging and FLOLA-Voronoi.}\label{fig:ishigamivariancetotal}
\end{figure}

In variance based sensitivity analysis, it is insufficient to only investigate the main effect sensitivity indices as strong interaction effects between the inputs may be present. This is the case for the Ishigami function of which the total sensitivity indices are shown in Figure~\ref{fig:ishigamitotalindices}. To accurately determine these total indices using the proposed algorithm, it is sufficient to only use variance based sensitivity cross-validation instead of the total variance based sensitivity cross-validation. Figure~\ref{fig:ishigamisensitivycross-validationvariancetotal} shows the variance based total sensitivity cross-validation measure which has no significant difference when compared to the variance based sensitivity cross-validation in Figure~\ref{fig:ishigamisensitivycross-validation}.

\begin{figure}[htb]
    \centering
    \begin{subfigure}[h]{0.48\textwidth}
        \includegraphics[width=\textwidth]{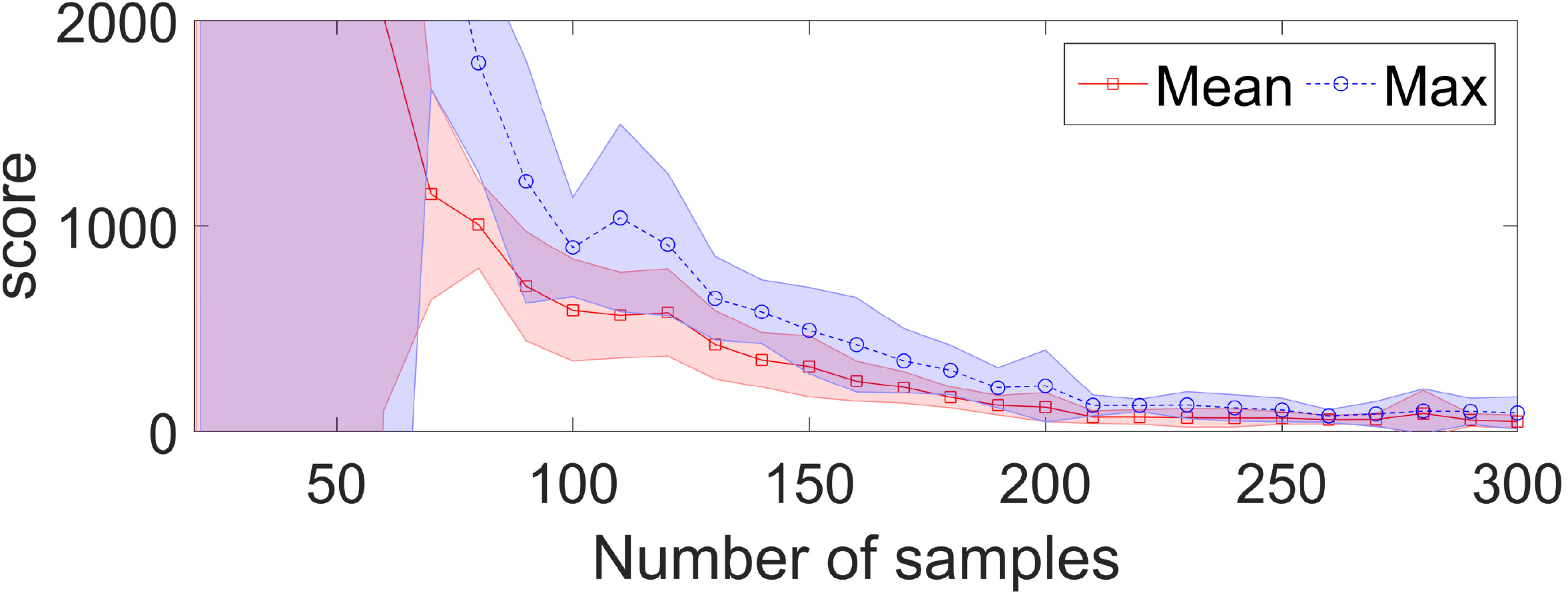}
        \caption{Sensitivity cross-validation.}
        \label{fig:ishigamisensitivitycross-validationderivative}
    \end{subfigure}
	~
    \begin{subfigure}[h]{0.48\textwidth}
        \includegraphics[width=\textwidth]{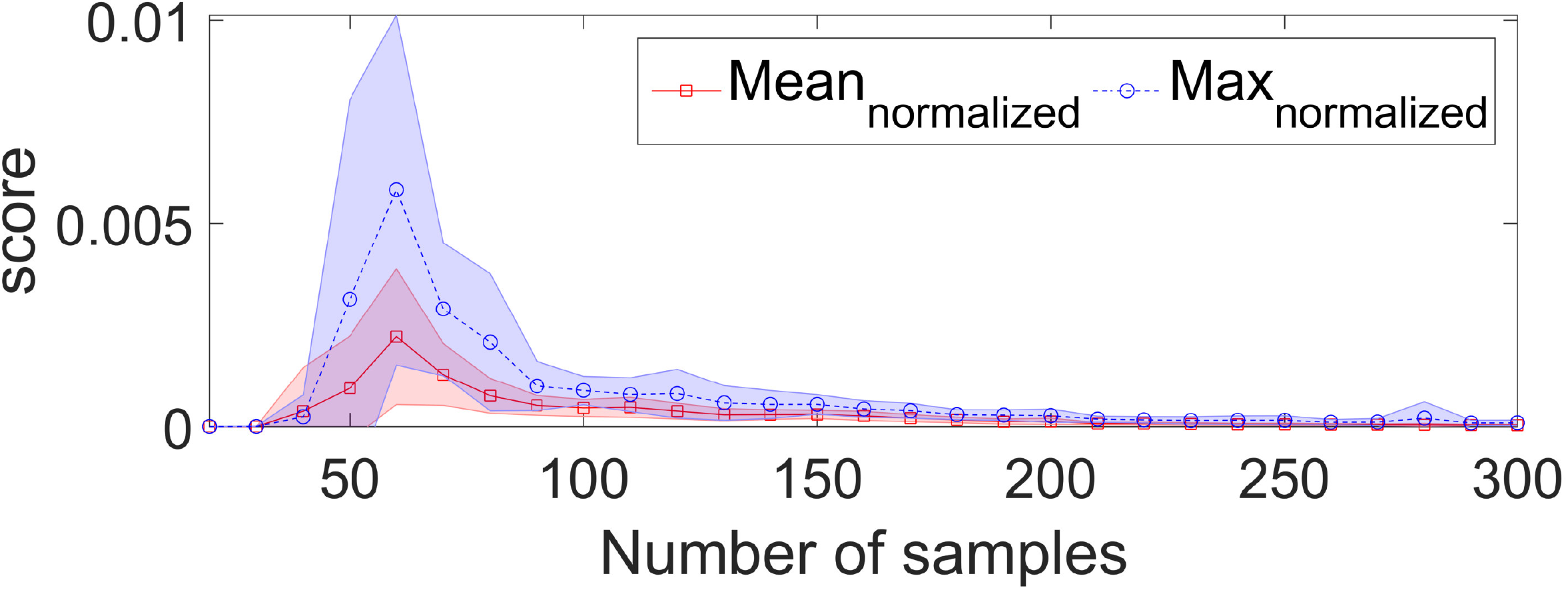}
        \caption{Normalized sensitivity cross-validation.}
        \label{fig:ishigamisensitivitycross-validationderivativenorm}
    \end{subfigure}
    
    \begin{subfigure}[h]{0.48\textwidth}
        \includegraphics[width=\textwidth]{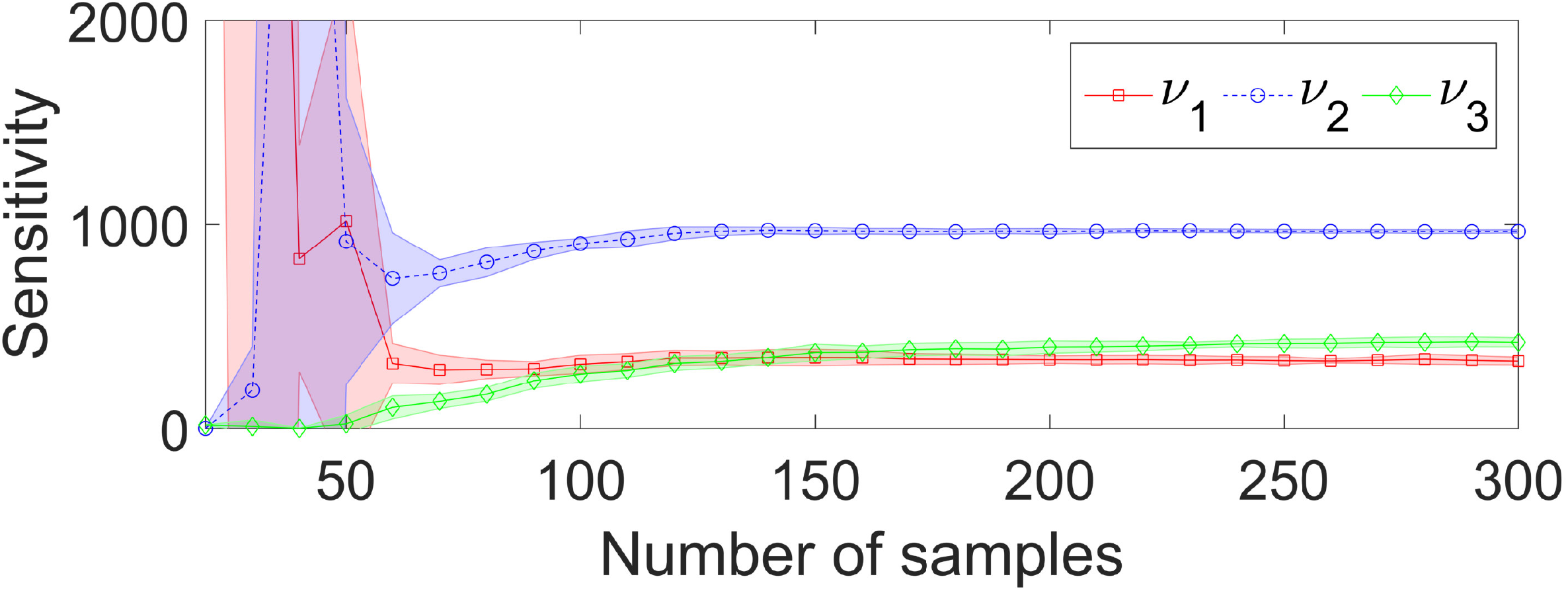}
        \caption{Sensitivity indices.}
        \label{fig:ishigamiderivativeindices}
    \end{subfigure}
    \caption{Derivative based sensitivity cross-validation with corresponding sensitivity indices for Ishigami with Kriging and FLOLA-Voronoi.}\label{fig:ishigamiderivative}
\end{figure}

The derivative based indices for the Ishigami experiment are shown in Figure~\ref{fig:ishigamiderivativeindices}. The derivative based sensitivity cross-validation measures are shown in Figure~\ref{fig:ishigamisensitivitycross-validationderivative}. As these indices reach large values, only the relevant part of the graph is shown in the figure. These large values are difficult to interpret as they are dependent on the indices. It is not possible to accurately determine a threshold for the sensitivity cross-validation without knowing the derivative based indices beforehand. The version of the sensitivity cross-validation measures in which the indices were first normalized is shown in Figure~\ref{fig:ishigamisensitivitycross-validationderivativenorm}. These measures can now be interpreted and a threshold can be determined.
Derivative based indices for the Ishigami function do not have the same ordering as the variance based total indices even though there is a link between them formulated by Equation~(\ref{eq:dgsmvstotalsobol}). This is due to the Ishigami function, which is highly non-linear, which influences the performance of derivative based sensitivity indices.

To analyze the accuracy of the calculated sensitivity indices, Table~\ref{table:ishigamicomparisonexact} shows a comparison between the experimentally determined indices and the exact calculated indices. These results were gathered with a final variance sensitivity cross-validation score of $2.4496\times10^{-6}$, total variance sensitivity cross-validation score of $2.6112\times10^{-6}$ and derivative sensitivity cross-validation score of $3.6477\times10^{-5}$.

\begin{table}[htb]
\centering
\caption{Comparison of experimentally calculated sensitivity indices with exact values for Ishigami with Kriging and FLOLA-Voronoi.}
\label{table:ishigamicomparisonexact}
\begin{tabular}{l|lll|lll}
\hline
Index & 1 & 2 & 3 & 1 & 2 & 3 \\ \hline
Type & \multicolumn{3}{c|}{\textbf{Experimental}} & \multicolumn{3}{c}{\textbf{Exact}} \\ \hline
\textbf{Variance} & 0.3152 & 0.4393 & 0.0002 & 0.3139 & 0.4424 & 0 \\
\textbf{Total-variance} & 0.5603 & 0.4397 & 0.2453 & 0.5576 & 0.4424 & 0.2437 \\
\textbf{Derivative} & 329.3 & 965.6 & 420.5 & 304.8 & 967.2 & 433.8 \\ \hline
\end{tabular}
\end{table}

To compare the performance of other metamodels, Table~\ref{table:ishigamicomparisonmetamodels} shows the experimentally determined sensitivity indices for the Ishigami function. For GP, the final sensitivity cross-validation scores were $2.2758\times10^{-6}$ for variance based, $2.4638\times10^{-6}$ for total variance based and $3.0642\times10^{-5}$ for derivative based. For LS-SVM, the final sensitivity cross-validation scores were $3.0547\times10^{-6}$ for variance based, $2.5661\times10^{-6}$ for total variance based and $1.7899\times10^{-5}$ for derivative based. The different metamodels perform similarly.

\begin{table}[htb]
\centering
\caption{Comparison of experimentally calculated sensitivity indices for GP and LS-SVM metamodels with FLOLA-Voronoi.}
\label{table:ishigamicomparisonmetamodels}
\begin{tabular}{l|lll|lll}
\hline
Index                   & 1         & 2         & 3        & 1          & 2          & 3          \\ \hline
Type                    & \multicolumn{3}{c|}{\textbf{GP}} & \multicolumn{3}{c}{\textbf{LS-SVM}} \\ \hline
\textbf{Variance}       & 0.3157    & 0.4393    & 0.0001   & 0.3066     & 0.4546     & 0.0000     \\
\textbf{Total-variance} & 0.5605    & 0.4397    & 0.2449   & 0.5453     & 0.4552     & 0.2387     \\
\textbf{Derivative}     & 326.3     & 965.9     & 405.9    & 301.1      & 1010.7     & 406.6      \\ \hline
\end{tabular}
\end{table}

\begin{figure}[H]
	\begin{subfigure}[h]{0.48\textwidth}
        \includegraphics[width=\textwidth]{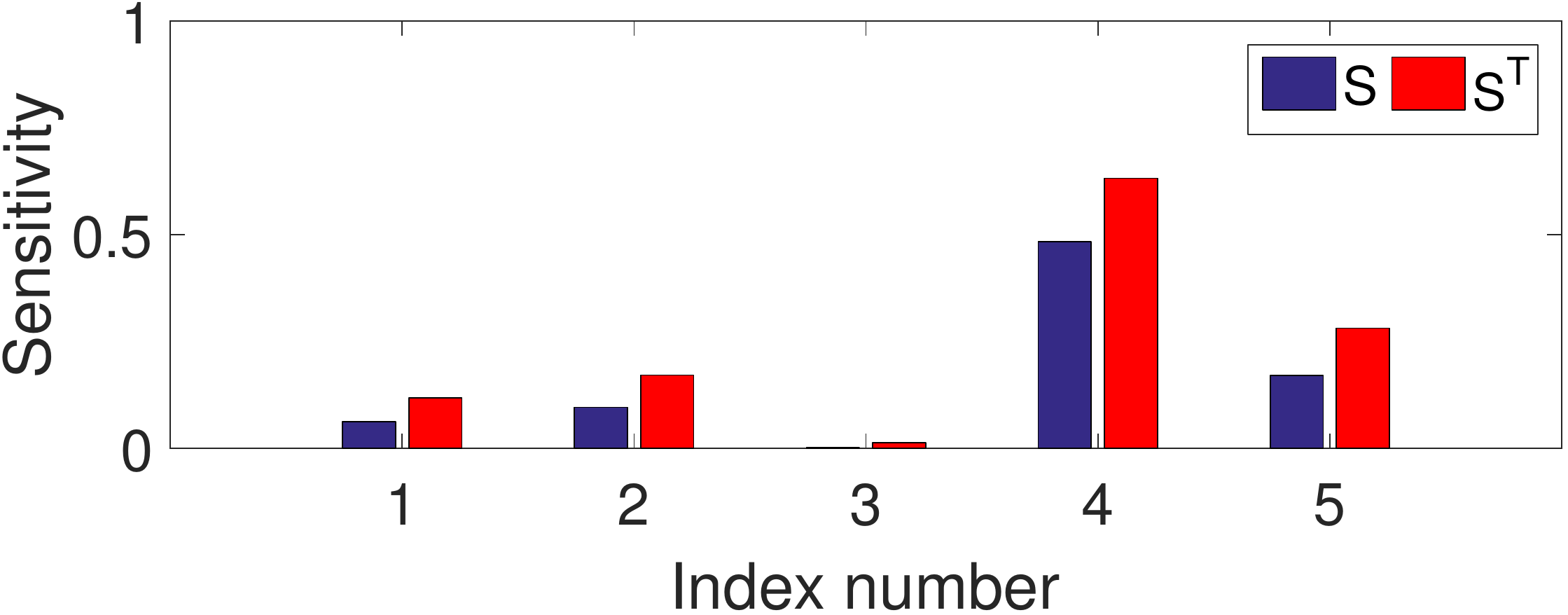}
        \caption{Variance based.}
        \label{fig:aaltobreakvarianceindices}
    \end{subfigure}
    ~
   	\begin{subfigure}[h]{0.48\textwidth}
        \includegraphics[width=\textwidth]{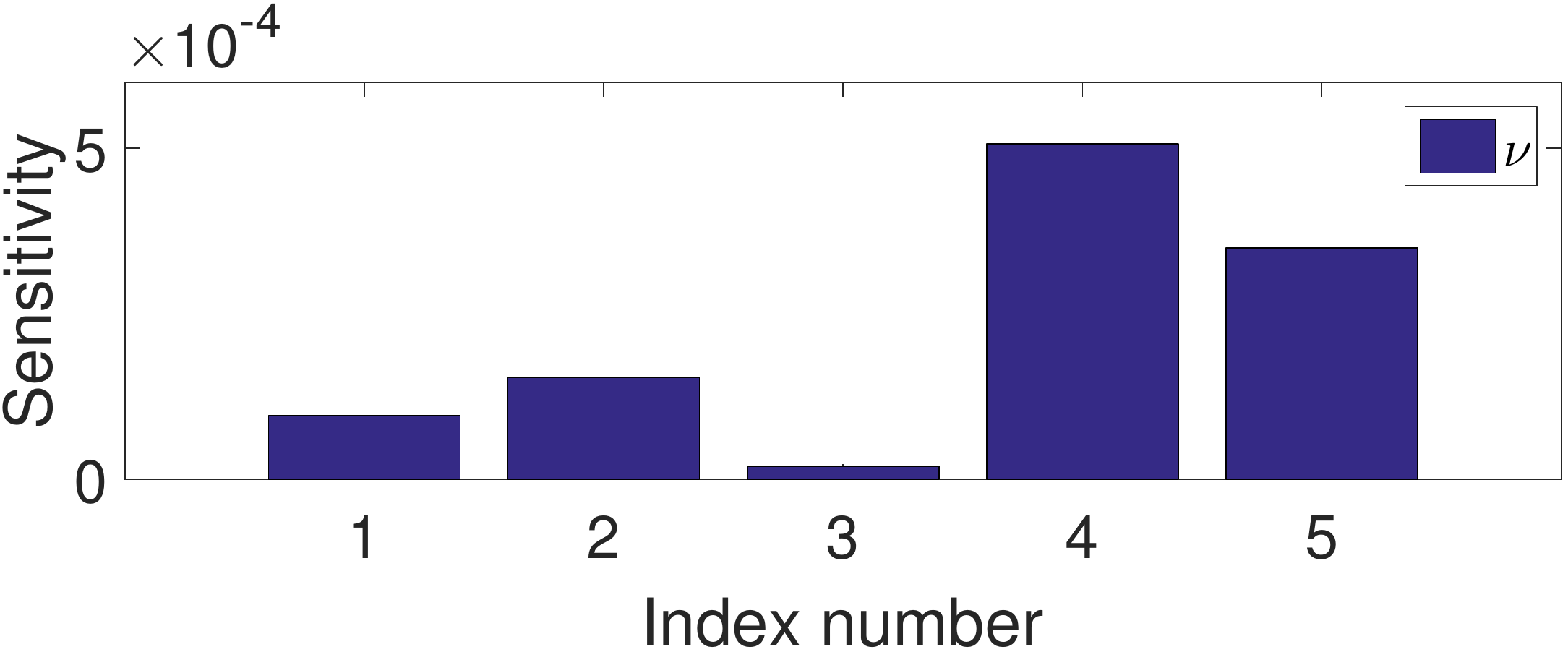}
        \caption{Derivative based.}
        \label{fig:aaltobreakderivativeindices}
    \end{subfigure}
    \caption{Sensitivity indices for AaltoBrake application with Kriging and FLOLA-Voronoi.}\label{fig:aaltobreakindices}
\end{figure}

Finally, we also present the performance of the proposed algorithm on the real-world AaltoBrake application. Figure~\ref{fig:aaltobreakindices} shows the sensitivity indices for the AaltoBrake application. From this, we learn that the mass of the brake system has no main impact on the output.

\section{CONCLUSION}
\label{sec:conclusion}
In this paper we proposed a complete global sensitivity analysis scheme for analysis of expensive black-box simulators. The use of a sequential design approach for metamodeling allows an efficient determination of the indices whilst the analytic derivations allow for an increased accuracy. Furthermore, the defined stopping criteria allow the analyst to stop the expensive evaluations when the desired predetermined accuracy threshold has been reached, depending on the application. This prevents doing unnecessary extra simulator evaluations, drastically reducing the total execution time. The results show a good performance for all tested metamodels and sequential designs. The techniques described are general and can be applied to other metamodels such as neural networks which is part of ongoing research. Using these results, researchers and designers can get better and faster insights into expensive black-box problems.

\section*{ACKNOWLEDGMENTS}
This research has (partially) been funded by the Inter university Attraction Poles Programme BESTCOM initiated by the Belgian Science Policy Office. Ivo Couckuyt is a post-doctoral research fellow of FWO-Vlaanderen.

\bibliographystyle{wsc}
\bibliography{paperbibliography}
\section*{AUTHOR BIOGRAPHIES}

\noindent {\bf TOM VAN STEENKISTE} received his M.Sc. degree in computer science engineering from Ghent University in July 2016. As of August 2016, he started as a PhD student at Ghent University within  the Internet Based Communication Networks (IBCN) research group. His interests are in simulations, metamodeling, machine learning and optimizations. His email address is \href{mailto:tdvsteen.vansteenkiste@ugent.be}{tdvsteen.vansteenkiste@ugent.be}.
\\

\noindent {\bf JOACHIM VAN DER HERTEN} 
received his M.Sc. degree in Computer Science in 2013, from the University of Antwerp, Belgium. Starting from August 2013 he is active as a PhD student in the research group Internet Based Communication Networks and Services (IBCN) at Ghent University. His research interests include surrogate modeling methods with sequential design including surrogate based optimization, active learning, optimal learning and machine learning. His e-mail address is \href{mailto:joachim.vanderherten@ugent.be}{joachim.vanderherten@ugent.be}.
\\

\noindent {\bf IVO COUCKUYT} is a postdoctoral research fellow at Ghent University - iMinds. He received the Master degree in Computer Science in 2007 from the University of Antwerp. Since then, he worked as a PhD student in the IBCN research group of the Department of Information Technology (INTEC) in the Faculty of Engineering at Ghent University, where he obtained the PhD degree in Science in 2013. His research is mainly focused on global and local surrogate modeling (metamodeling) and its application to solve real world problems, optimization of expensive functions,  evolutionary computing and machine learning methods. He is also the lead developer and project manager of the Surrogate Modeling (SUMO) Toolbox and the ooDACE Toolbox. His e-mail address is \href{mailto:ivo.couckuyt@ugent.be}{ivo.couckuyt@ugent.be}.
\\

\noindent {\bf TOM DHAENE} is full professor at the Department of Information Technology (INTEC) of Ghent University. He received the PhD degree in electrical engineering from Ghent University, Belgium, in 1993.  In September 2000, he joined the Department of Mathematics and Computer Science of the University of Antwerp as a Professor. Since October 2007, he has been a Full Professor with the Department of Information Technology, Ghent University. He is also affiliated with iMinds. As author or co-author, he has contributed to more than 350 peer-reviewed papers and abstracts in international conference proceedings, journals and books about computational science and numerical analysis and engineering. He is the holder of five U.S. patents. His e-mail
address is \href{mailto:tom.dhaene@ugent.be}{tom.dhaene@ugent.be}.
\\
\end{document}